\documentclass[11pt]{article}
\newif\iffigs\figstrue

\usepackage{amsmath}
\usepackage{amssymb}

\DeclareMathAlphabet{\mathpzc}{OT1}{pzc}{m}{it}

 \csname
@addtoreset\endcsname{equation}{section}

\def\gz0{\gamma^{0}}

\def\be{\begin{equation}}
\def\ee{\end{equation}}
\def\bea{\begin{eqnarray}}
\def\eea{\end{eqnarray}}
\def\ba{\begin{array}}
\def\ea{\end{array}}
\def\bec{\begin{center}}
\def\ec{\end{center}}
\def\ba{\begin{align}}
\def\ena{\end{align}}

\def\12{\frac{1}{2}}

\usepackage{slashed}

\usepackage{float}

\usepackage{booktabs}

\usepackage[mathscr]{euscript}

\usepackage{color}
\usepackage{graphicx}
\usepackage{epsf}
\usepackage{graphicx,epsfig}
\usepackage{amsmath}

\usepackage{multirow}

\usepackage{amsmath, amsthm, amssymb}

\usepackage{epsfig}
\usepackage{cite}
\usepackage{color,colordvi}

\newcounter{hran}

\makeatletter
\renewcommand\section{\@startsection {section}{1}{\z@}%
                               {-3.5ex \@plus -1ex \@minus -.2ex}%
                               {2.3ex \@plus.2ex}%
                               {\normalfont\large\bfseries}}
\makeatother

\vspace{0.5cm}

\renewcommand{\bea}{\begin{eqnarray}}
\renewcommand{\eea}{\end{eqnarray}}
\newcommand{\bi}{\begin{itemize}}
\newcommand{\ei}{\end{itemize}}

\def\beq{\begin{equation}}
\newcommand{\eeq}[1]{\label{#1}\end{equation}}

\begin{document}
\thispagestyle{empty}
\begin{flushright}
CERN-PH-TH-2015-193
\end{flushright}

\vspace{10pt}

\begin{center}


{\Large\sc Scale Invariant Volkov--Akulov Supergravity}\\


\vspace{25pt}
\begin{center}
{\sc S.~Ferrara${}^{\; a,b,c}$, M.~Porrati${}^{\; a,d}$~\footnote{On sabbatical at CERN-Ph-Th until September 1, 2015.} \ and \ A.~Sagnotti${}^{\; a,e}$~\footnote{On sabbatical at CERN-Ph-Th until September 30, 2015.}}\\[15pt]

{${}^a$\sl\small Th-Ph Department, CERN\\
CH - 1211 Geneva 23 \ SWITZERLAND \\[4pt]

${}^b$ INFN - Laboratori Nazionali di Frascati \\
Via Enrico Fermi 40, 00044 Frascati \ ITALY\\[4pt]

${}^c$ Department of Physics and Astronomy,
University of California \\ Los Angeles, CA 90095-1547 \ USA\\[4pt]
}

{${}^d$\sl\small CCPP, Department of Physics, NYU \\ 4 Washington Pl., New York NY 10003, USA}\vspace{4pt}

{${}^e$\sl\small
Scuola Normale Superiore and INFN\\
Piazza dei Cavalieri \ 7\\ 56126 Pisa \ ITALY \\[12pt]}

{\small \sl
sergio.ferrara@cern.ch, mp9@nyu.edu, sagnotti@sns.it}
\end{center}
\vspace{24pt}

\vspace{5pt} {\sc\large Abstract}\end{center}

\noindent A scale invariant Goldstino theory coupled to supergravity is obtained as a standard supergravity dual of a rigidly scale--invariant higher--curvature supergravity with a nilpotent chiral scalar curvature.
The bosonic part of this theory describes a massless scalaron and a massive axion in a de Sitter Universe.

\noindent

\vfill
\baselineskip=20pt
\noindent

\setcounter{page}{1}

\pagebreak

\newpage

\section{\sc \bf Introduction}

Motivated by single--field inflationary scenarios\cite{inflation}, several sgoldstinoless \cite{va_1,va_2,va_3,va_4} supergravity extensions of inflationary models were recently considered \cite{sgoldstinoless_0, sgoldstinoless_1,sgoldstinoless_2,sgoldstinoless_3,sgoldstinoless_4,sgoldstinoless_5} (for a recent review see \cite{fs0615}). Interestingly enough, in \cite{sgoldstinoless_1,sgoldstinoless_5} many of these models were linked to pure higher--derivative supergravity with a nilpotency constraint on the scalar curvature chiral superfield ${\cal R}$. These include the Volkov--Akulov--Starobinsky model\cite{sgoldstinoless_1} and the pure Volkov--Akulov theory coupled to supergravity \cite{sgoldstinoless_1}. Recently, the full component form of the latter theory was presented in \cite{va_comp1,va_comp2}

Along these lines, various authors considered $R^{\,2}$ theories of gravity \cite{r2_1} and their supergravity embeddings \cite{r2_1,r2_2}, which possess a rigid scale invariance and naturally accommodate a de Sitter Universe. It is the aim of this note to give the sgoldstinoless version of these theories, which naturally combines an enhanced rigid scale invariance and a de Sitter geometry. This theory also emerges as a limiting case of the inflationary scenario.

\section{\sc \bf Scale--Invariant Nilpotent Supergravity}

The superspace action density of the scale--invariant theory that we consider~\footnote{We use throughout the conventions of \cite{sgoldstinoless_1}.},
\begin{equation}
{\cal A} \ = \ \left. \frac{{\cal R}\, {\cal {\overline R}}}{g^2} \right|_D
\ + \  \, \sigma \, {\cal R}^2 \, S_0 \bigg|_F   \ ,
\label{scale1}
\end{equation}
where $g$ is a dimensionless parameter, is invariant under the rigid scale transformations
\begin{equation}
{\cal R} \ \rightarrow \ {\cal R} \ , \ S_0 \ \rightarrow \ e^{\,-\,\lambda}\, S_0 \ , \ \sigma \ \rightarrow \ e^{\,\lambda}\, \sigma \ . \label{scale2}
\end{equation}
This theory is equivalent to the theory considered in \cite{r2_2}, supplemented with the nilpotency constraint
\begin{equation}
{\cal R}^{\,2} \ = \ 0 \ , \label{scale3}
\end{equation}
which is enforced by the chiral Lagrange multiplier $\sigma$ present in the second term of eq.~\eqref{scale1}.

Using manipulations similar to those originally introduced in \cite{cecotti}, we can now turn this model into a scale--invariant version of the Volkov--Akulov model coupled to standard supergravity. To this end, we first use the superspace identity
\begin{equation}
\sigma \, {\cal R}^2\, S_0 \ + \ h.c.  \bigg|_F \ = \ \left. \left(\sigma \, \frac{\cal R}{S_0}\ +\
{\overline \sigma} \, \frac{\cal {\overline R}}{{\overline S}_0}\right) S_0 \, {\overline S}_0 \right|_D \ + \ {\rm tot. \ deriv.} \ , \label{scale4}
\end{equation}
and then introduce two Lagrange chiral superfield multipliers $T$ and $S$ according to
\begin{equation}
{\cal A} \ = \ \left. \bigg(\, \sigma\, S\   + \ {\overline \sigma}\, {\overline S} \ +\
\frac{ S \,{\overline S}}{g^2} \bigg) S_0 \,{\overline S}_0 \right|_D \, - \,
\left. T \left(\frac{\cal R}{S_0}\,-\,S \right) S_0^3 \ + \ {\rm h.c. } \right|_F \ .
\label{scale45}
\end{equation}
The final result is the standard supergravity action density
\begin{equation}
{\cal A} \ = \ - \, \left. \bigg(\,T \,+\, {\overline T} \, {-} \, \sigma \,S \, {-} \, {\overline \sigma}\, {\overline S} \,-\,
\frac{S \, {\overline S}}{g^2}\bigg) S_0 \, {\overline S}_0 \right|_D  +\,
 T \, S \, S_0^3 \,+\, {\rm h.c. } \bigg|_F \ + \ {\rm tot. \ deriv.} \label{scale5}
\end{equation}
A final shift and a redefinition according to
\begin{equation}
T \ \rightarrow \ T \  +\  \sigma \, S \ , \qquad X \ = \ \frac{S}{g} \label{scale6}
\end{equation}
yield the standard supergravity action density
\be
{\cal A} \ = \  - \  (T \ + \ {\overline T} \ - \ X \, {\overline X}) \, S_0 \, {\overline S}_0 \bigg|_D \ + \  W(T,X) \, S_0^3 \ +\
{\rm h.c.} \bigg|_F  \ , \label{scale8}
\ee
where
\begin{equation}
W(T,X,\sigma) \ = \ g\,T\,X \ + \ g^2 \, \sigma \, X^2 \ .
\label{scale9}
\end{equation}
This is tantamount to the scale--invariant superpotential
\begin{equation}
W(T,X) \ = \ g\,T\,X \ ,
\label{scale10}
\end{equation}
where $X$ is subject to the nilpotency constraint
\begin{equation}
X^{\,2} \ = \ 0 \ ,
\label{scale11}
\end{equation}
so that $X$ describes the sgoldstinoless Volkov--Akulov multiplet \cite{va_1,va_2,va_3,va_4}. The corresponding bosonic Lagrangian,
\begin{equation}
{\cal L} \ = \ \frac{R}{2} \ - \ \frac{3}{(T \,+ \, {\overline T})^2} \ |\partial \,T|^2\  - \ g^{\,2}\, \frac{|T|^2}{3 ( T \,+ \,{\overline T})^2}   \ , \label{scale12}
\end{equation}
is a special case of the result displayed in \cite{sgoldstinoless_1}, so that it describes an $SU(1,1)/U(1)$ K\"ahlerian model of curvature -2/3 with a scale--invariant positive potential. As a result, in terms of the canonical variable
\be
T \ = \ e^{\,\phi{\sqrt \frac{2}{3}}} \ + \ i \, a \, \sqrt{\frac{2}{3}} \ , \ \label{scale13}
\ee
one finds
\begin{equation}
{\cal L} \ = \ \frac{R}{2} \ - \ \frac{1}{2}\ (\partial \phi)^2 \ - \ \frac{1}{2} \ e^{\,-\,2\, \phi \,{\sqrt \frac{2}{3}}}\
(\partial a)^2 \ - \ \frac{g^{\,2}}{12} \ - \ \frac{g^{\,2}}{18} \ e^{- 2\,\phi
{\sqrt \frac{2}{3}}} \ a^2 \ . \label{scale14}
\end{equation}

Note that in the Einstein frame the metric is inert under the scale transformation corresponding to eq.~(\ref{scale2}), while
\begin{equation}
\phi \ \rightarrow \ \phi \ + \ \gamma \ , \quad a \ \rightarrow \ e^{\gamma\, \sqrt{\frac{2}{3}}} \ a \ . \label{scale15}
\end{equation}
%
\section{\sc \bf de Sitter Vacuum Geometry}

Since $a$ is stabilized at zero, this model results in a de Sitter vacuum geometry, with a corresponding scale--invariant realization of supersymmetry breaking induced by the non--linear sgoldstinoless multiplet. The supersymmetry breaking scale $M_s^2$ is
\begin{equation}
M_s^2 \ = \ \frac{g}{2\,\sqrt{3}} \ M_{Planck}^{\,2} \ ,
\label{scale16}
\end{equation}
up to a conventional numerical factor.
Eq.~\eqref{scale8} describes the minimal supergravity model that embodies a scale--invariant goldstino interaction and leads unavoidably to a de Sitter geometry. This model involves a single dimensionless parameter $g$, which determines its \emph{positive} vacuum energy according to
\begin{equation}
V \ = \ \frac{g^{\,2}}{12} \ M_{Planck}^{\,4}\ .
\label{scale18}
\end{equation}
In contrast, the Volkov--Akulov model coupled to supergravity, depends on the two parameters $f$ and $W_0$, and consequently leads to a vacuum energy \cite{supergravity} \cite{dz}\cite{cfgv} \cite{sgoldstinoless_1}\cite{va_comp1}
\begin{equation}
V \ = \ \frac{1}{3} \ \left| f \right|^2 \ - \ 3\, \left|W_0 \right|^2 \label{scale17}
\end{equation}
of arbitrary sign.

\vskip.3in
\noindent
{\bf Acknowledgments}

\noindent We would like to thank I.~Antoniadis, E.~Dudas, R.~Kallosh, A.~Kehagias, A.~Linde and A.~Van Proeyen for collaboration on related issues. S.F. is supported in part by INFN (I.S. GSS). M.P. is supported in part by NSF grant PHY--1316452, while A.S. is supported in part by Scuola Normale and by INFN (I.S. Stefi).

\end{document}